# A Secure Design Pattern Approach Toward Tackling Lateral-Injection Attacks

Chidera Biringa
University of Massachusetts Dartmouth
Dartmouth, USA
cbiringa@umassd.edu

Gökhan Kul
University of Massachusetts Dartmouth
Dartmouth, USA
gkul@umassd.edu

*Abstract*—Software weaknesses that create attack surfaces for adversarial exploits, such as lateral SQL injection (LSQLi) attacks, are usually introduced during the design phase of software development. Security design patterns are sometimes applied to tackle these weaknesses. However, due to the stealthy nature of lateral-based attacks, employing traditional security patterns to address these threats is insufficient. Hence, we present SEAL, a secure design that extrapolates architectural, design, and implementation abstraction levels to delegate security strategies toward tackling LSQLi attacks. We evaluated SEAL using case study software, where we assumed the role of an adversary and injected several attack vectors tasked with compromising the confidentiality and integrity of its database. Our evaluation of SEAL demonstrated its capacity to address LSQLi attacks.

## I. INTRODUCTION

SQL injection attacks constitute a specialized set of attacks where an adversary injects malicious inputs to compromise the security of software or network [1]. An NTT report [2] presented in a case study that SQL injection attacks alone cost up to $196,000 to an anonymized national bank, which emphasizes the importance of robust defense systems. Common Weakness Enumeration (CWE) regularly spotlights various injection attack types on the top 25 most dangerous software weaknesses [3]. Common targets of injection attacks are software that allows the insertion of inputs, such as web applications. In 2021, The Open Web Application Security Project (OWASP) ranked injection attacks top 3 significant threats to the security of the web [4]. These attacks violate confidentiality, integrity, availability, and traceability (CIAT) security concerns. In severe cases, they can potentially lead to the total unavailability of critical services. Lateral SQL Injection (LSQLi) attacks are derived from injection attacks where an adversary conducts exploits in fragments through time [5]. LSQLi differentiates itself from SQLi attacks by persistently adopting multiple attack strategies to compromise software security. After obtaining initial access, an adversary employs several malicious mechanisms to stealthily and progressively traverse through the system gaining elevated privileges and trust levels. Secure design patterns have been used to prevent the accidental or intentional introduction of software weaknesses during the design phase of software development [6], [7]. However, to the best of our knowledge, no research has investigated tackling lateral-SQLi attacks from the design level. The study of SQLi and lateral-SQLi attacks are comprehensive and reliable solutions proposed [8], and language-dependent measures such as prepared statements are adopted [9]. Recently, machine learning-based approaches [10] have produced good results in this regard. The goal of this paper is to explore the application of a secure design pattern in addressing lateral SQLi attacks and not to propose concrete secure methods and algorithms that prevent lateral SQLi attacks from occurring. Hence, we present a **S**ecure D**E**sign pattern **A**pproach towards tackling **L**ateral-injection attacks – *SEAL*. SEAL is a secure design pattern that decomposes user and security level features into independent but collaborative components to tackle lateral-based in-band SQL injection attacks. We split SEAL into three zones (i) Injection Zone (IZ), (ii) Sensitive Zone (SenZ), and (iii) Security Zone (SecZ). The IZ models a user interaction component through which a potential adversary injects single or multiple attack vectors to compromise software security. The SenZ is where we store and manage sensitive data, such as user credentials and authorization privileges. SecZ is the core component of SEAL, responsible for accommodating secure algorithms to tackle lateral-based SQLi, which provides the insertion and deployment of concrete security algorithms.

**Outline.** In Section II, we describe the necessary background for this work. Sections III and IV details proposed design and threat model analysis. Section V evaluates SEAL and Section VI concludes this paper.

## II. BACKGROUND

**Lateral SQLi Attacks: Inband Variant.** Lateral in-band SQLi attacks are a variant of SQLi attacks. In this case, an adversary executes a lateral-augmented attack using traditional entry points. Technically-adept adversaries typically employ several attack vectors and strategies when attacking a system, as in the case of Advanced Persistent Threats (APT) [11]. Thus, it is pertinent to have secure software systems that are reactive to the evolving behavior of attack strategies. For example, Figure 2 is a lateral SQLi attack tree that models adversarial attacks comprising a collection of SQLi methods. In this scenario, an adversary utilizes a multifaceted strategy involving the persistent injection and substitution of payloads until the attack is advanced and compromise successful.

We demonstrate SEAL using an in-band SQLi attack. An in-band is the most typical of SQLi attacks. It describes an attack that uses the usual communication medium as exploits point-of-entry and result generation [12]. Common types of in-band SQLi include: UPDATE, ERROR, and UNION-based attacks [12]. UPDATE-based attacks exploit the UPDATE SQL operator to persist erroneous data in the database. ERROR-based attacks utilize improperly thrown errors by the database server to derive sensitive information about the structure and schema of the database.

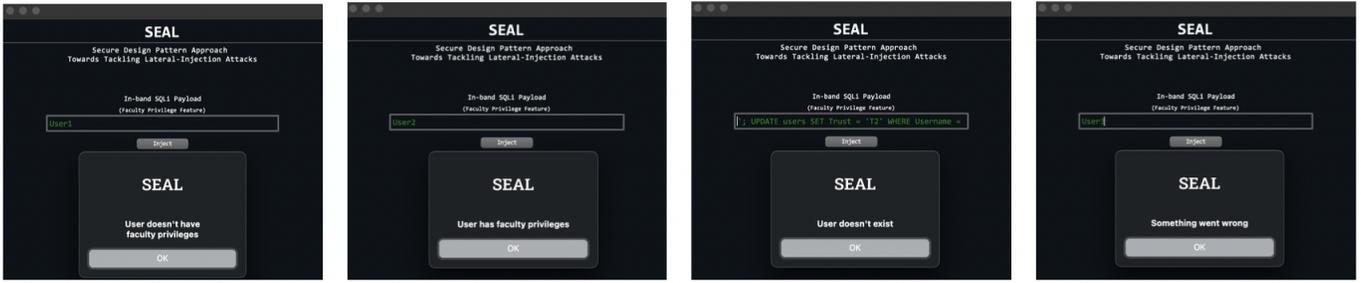

Fig. 1: A simplified demonstration of lateral SQLi exploits and secure strategy delegation using SEAL.

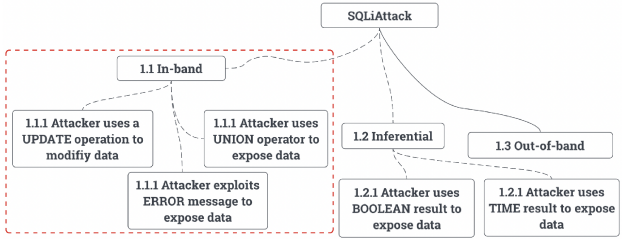

Fig. 2: SQLi Attack Tree. Dotted and dashed lines indicate that the attack is highly probable and improbable. Components in the dotted red box are in-band SQLi attack variants.

### III. SEAL: Secure Design Pattern

**SEAL's Overall Strategy.** The primary objective of SEAL is to delegate and substitute appropriate secure modules to tackle lateral-SQLi. We describe its modules and the interaction between them from the inception of an injected attack vector to the selection of security modules. We focus on the UPDATE and ERROR-based variants of SQLi. In Figure 3, we display the structure of the design pattern. In designing SEAL, we consider three major factors: (i) the design should be generalized to all high-level programming languages, (ii) the design should facilitate the decoupling of user and security functionalities, and (iii) the design endpoints should be sufficiently flexible to permit the easy insertion of concrete security algorithms. SEAL comprises three design subsystems and one architectural-level design. On the architectural level, we distrustfully decompose processes into Injection (IZ), Sensitive (SenZ), and Secure Zones (SecZ). The IZ comprises a user interaction subsystem representing the communication medium and a single access injection point for an adversary to insert an attack vector or a collection of attack vectors. We utilize a graphical user interface to facilitate user interaction. SenZ represents the data we want to protect from malicious activities. In our case, this is the Sensitive database and its tables: User and Authorization. Finally, SecZ is the core component of SEAL and is responsible for selecting relevant security strategies to tackle these attacks. We detail independent modules that comprise SecZ.

**Secure Zone: Secure Strategy Delegation Process.** SEAL enhances security protection by providing a clear bifurcation of security strategies that address lateral in-band SQLi attacks and user-level functionality. The secure modules are augmented to dynamically create objects in response to streams of attack vectors from an adversary. Given an injected attack vector through the IZ, the Delegator module acts as a master controller, which receives input from the IZ and calls the InputValidation module that performs a light input validation, validating the data for type. Post-validation, the payload is sent to the ThreatHandler module, which detects whether or not it is an UPDATE-based or ERROR-based injection attack through the ThreatDetector interface. Next, the Delegator calls the FactoryHandler module, which handles the creation of concrete security strategies through the SecureFactory interface. Finally, the SQLiContext modules take a factory argument and execute a `delegate_strategy()` method, which calls the secure functionality from the SensitiveZone to address the injected attack vector.

### IV. Threat Model: Demonstrating Exploits and Secure Strategy Delegation

**The Anatomy of an In-band SQLi.** We designed in-band SQLi exploits (E) to demonstrate the utility of our proposed design in tackling injected attacks by permitting the integration of concrete secure strategies. These attack vectors were injected into our case study software and compromised its security. Finally, we describe how these exploits are addressed using SEAL.

**$E_1$: UPDATE-based In-band SQLi.** The UPDATE-based exploit maliciously modifies and persists database content. We model the attack as the acquisition of initial access by the adversary. Embedded in SEAL's SenZ is an SQLite relational database. The database consists of Users and Authorization tables. Users table consists of four non-identity columns: (i) Username, (ii) Student, (iii) Faculty, and (iv) Trust, while the Authorization table consists of 2 non-identity columns: (i) Trust and (ii) Privilege. The Username column in the Users table represents a valid student or faculty member. Student and Faculty columns denote whether or not a user is a student or a faculty member. If the user is a student, the column value is `True` and `False` for `Faculty`. If the user is a faculty member, the column value is `True` and `False` for student. The `Privilege` column in the `Authorization` table denotes the authorization privileges of users. `Trust` column represents a component in threat modeling-application decomposition process [13] where we assigned "trust levels" privileges to valid users. In this context, T1 implies that a student user only has authorization to `View Grades`, and T2 implies a faculty user only has the authorization to `Enter Grades`. An adversary with knowledge of this schema can design SQLi attacks to compromise the confidentiality and integrity of the database.

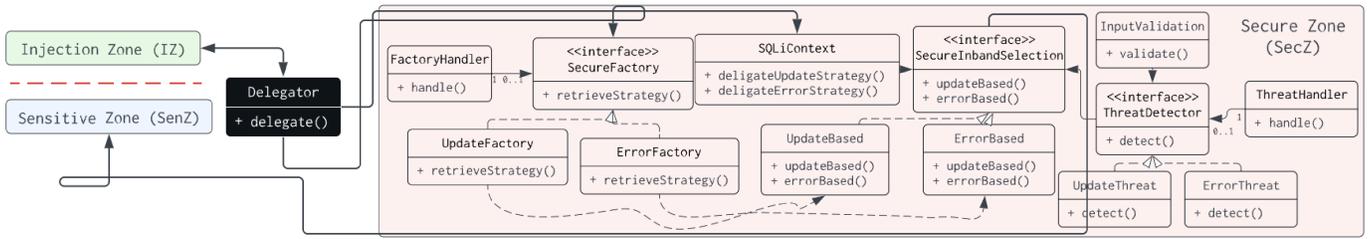

Fig. 3: SEAL: Secure Design Pattern. A trust boundary [13] exists between the IZ and SenZ. Hence, all data requests from the IZ must pass through the SecZ secure strategy delegation. SenZ is only accessible through SecZ.

**Root Cause Analysis.** Listing 1 displays the result of an executed query that obtains the authorization privileges assigned for all users.

Listing 1: User Authorization Privilege
```
out: ('User1', 'View Grades')
     ('User2', 'Enter Grades')
```

"User1" has sole authorization to "View Grades" while "User2" is only authorized to "Edit Grades." Given that user privileges are contingent on trust levels (T1, T2), an attacker can exploit this by injecting malicious queries to assign T2 authorization privileges to a student user.

Listing 2: A Simple SQLi Exploit
```
"'; UPDATE users SET Trust = 'T2' WHERE Username = 'User1';
    SELECT 1; --"
```

Listing 3: Vulnerable Faculty Privilege Function
```
1 def has_entergrades(username):
2   with connect(self.sensitive) as conn:
3     cur = conn.cursor()
4     cur.executescript("SELECT Trust FROM users WHERE Username
         = '%s'" %username)
5     trust_level = cur.fetchone()
6     if trust_level is None:
7       return ValueError("User doesn't exist!").__repr__()
8     return True if trust_level[0] == "T2" else False
```

Listing 4: Secure Faculty Privilege Status
```
cur.execute("SELECT Trust FROM users WHERE username = ?",
    (username,))
```

In Listing 2, we display a simple SQLi exploit. We feed the attack vector to the faculty privilege verification feature shown in listing 3. It has a `username` input parameter, which takes a name argument and returns a faculty status, where $\{1 \in T2\}$ and $\{0 \notin T2\}$. In this case, listing 3, line 4 is the specific vulnerable code snippet we will be exploiting. We pass the SQLi attack vector as a argument to the `has_entergrades()` method. Post-exploit, we execute the user authorization privilege query in listing 1, and the result in listing 5 shows that "User1", a student user has the authorization privilege to "Enter Grades."

Listing 5: Post-Exploit User Authorization Privilege
```
out: ('User1', 'Enter Grades')
     ('User2', 'Enter Grades')
```

The consequence of this exploit is apparent. The attacker has access to modify the student's course grade after being entered by a faculty member. To achieve a fundamental comprehension of how and why the SQLi exploits work, we investigate the independent component of the query. In Listing 2, `(';)` nullifies the original query, succeeding statement: (**UPDATE** users **SET** Trust = 'T2' **WHERE** Username = ↪ 'User1';) performs an update operation where "User1" is assigned trust level T2. Finally, (**SELECT** 1; `--)` always returns true and invalidates subsequent statements using `(--)` single-line comment.

We have detailed how an adversary can conduct an in-band SQLi exploit to compromise the confidentiality and integrity of a database system. We have also described the consequences associated with this attack. Next, we discuss concrete secure strategies plugged into SEAL to tackle the above-stated vulnerability. In Listing 4, we have adopted secure development practices to rewrite the vulnerable code snippet in line 4 of the `has_entergrades()` function. Inspecting the code, we see that we have eliminated the `executescript()` method — which supports the execution of multiple queries consecutively, creating an attack surface vulnerable to injection — and replaced it with `execute()`. Next, we use secure-query parameters rather than passing interpolating parameters, which parse arguments directly to the database without input validation or sanitization. We want to note that described defense strategies in this work are solely demonstrative. It shows how SEAL — a secure design pattern can be utilized and extended. It doesn't constitute a significant depth of SQLi security measures, which is outside the scope of our work.

**E$_2$: ERROR-based In-band SQLi.** It is an oversight to assume that the `has_entergrades()` functionality is entirely secure. In the course of addressing the UPDATE-based SQLi attack, we have exposed secret information to the adversary violating the information hiding modularization principle [14] and results in the exposure of sensitive information (CWE-200) [3] . Listing 3, line 7 returns a (**ValueError**("User

`doesn't exist!"))` stack trace error if the entered username is not in the database. The above-stated exposure might seem inconsequential to a less skilled adversary but not to an advanced adversary. For example, the `has_entergrades()` function checks whether or not supplied username has the appropriate trust levels to enter student grades and returns either a `True` or `False` status. This difference enables an adversary to comprehend the state of the feature and aid in discovering the authorization privileges of all users by trying permutations of user input until the correct user is retrieved. Hence, it makes it easier for an attacker to obtain some necessary authentication credentials. While this information may be helpful to a user, it is also useful to an attacker. To address this, we enumerate a whitelist [15] of valid users in the database, where the request fails securely if the injected username doesn't match the criteria. In practice, we change code in line 7 to (`return`), which handles the error by totally obfuscating the returned error message.

## V. EVALUATION

We evaluate SEAL's feasibility using four cases.

**Case 1: `User1` Benign Injection.** In this case, we enter a "User1" into the `entry field` of SEAL's IZ and click `Inject`. Given that (User1 ∈ Student), (Student ∈ Trust) and (T1 ∈ Trust), where (T1 ⊃ "View Grades" ∃ User1). The software returns a response message indicating that the injected input doesn't have `faculty` authorization privileges.

**Case 2: `User2` Benign Injection.** In this case, we enter a "User2" into the `entry field` of SEAL's IZ and click `Inject`. This case is the inverse of the first case, (User2 ∈ Faculty) with T2 trust levels, where (T2 ⊃ "Enter Grades" ∃ User2). The software returns a response message indicating that the injected user is a faculty member with authorization privileges.

**Case 3: UPDATE-based Malicious Injection.** In this case, we enter the listing 2 UPDATE-based attack vector, exhaustively described in Section IV in the `entry field` of SEAL's IZ, and click `inject`. The `Delegator` module receives the attack vector and calls the `InputValidation` module to validate the input, and passes validated data to the `ThreatHandler` module which detects an UPDATE-based threat and calls the `FactoryHandler` module, which creates an UPDATE-based secure factory. Finally, the `Delegator` passes the above-stated factory as an argument to `SQLiContext`'s parameter, which calls the `delegate_update_strategy()` with the injected attack vector as an argument, concluding the tackling of UPDATE-based attack returning a "User doesn't exist" error message.

**Case 4: ERROR-based Malicious Injection.** In this case, the adversary is unsuccessful with the initial UPDATE-based attack and laterally proceeds to intentionally inject invalid inputs to exploit the error generation process of the database server to extract secret information about the `SensitiveZone`. The delegation process is similar to the third case, only with a different secure strategy. SEAL detects the threat pattern of the injected payload and delegates `ERROR-based` security strategy to address the attack, which includes the total obfuscation of error and returning a "Something went wrong" message. Our evaluation showed that SEAL is practical in selecting security strategies to address lateral in-band SQLi attacks. **The source code of SEAL and our case study software are publicly available: `https://github.com/biringaChi/SEAL`.**

## VI. CONCLUSION

Secure software design has proven reliable in providing standardized security strategies for common recurring software weaknesses and vulnerabilities. SEAL is a design pattern with characteristics that model secure interaction between sensitive and non-sensitive data in the software. To the best of our knowledge, there isn't an exploration and implementation of this design pattern to tackle lateral-augmented in-band SQL injection.

**Acknowledgments.** We want to thank Yi Liu, Ph.D. for her review of this paper.